\documentclass{iau}
 \usepackage{graphicx} 

\title[IAUS291.~~On the Glitch Evolution of Pulsars] 
{On the Glitch Evolution of Pulsars} 

\author[Urama, Joshi \& Chukwude]  
{J.O. Urama$^1$, B.C. Joshi$^2$
 \and A.E. Chukwude$^1$}

\affiliation{$^1$Department of Physics \& Astronomy,
   University of Nigeria, Nsukka \\ email: {\tt johnson@hartrao.ac.za}(JOU); {\tt aus\_chukwude@yahoo.com}(AEC) \\[\affilskip]
$^2$National Centre for Radio Astrophysics, Post Bag 3, Ganeshkhind, Pune 411 007, India \\email: {\tt bcj@ncra.tifr.res.in}}

\pubyear{2012}
\volume{291}  
\jname{\mbox{Neutron Stars and Pulsars: Challenges and Opportunities after 80 years}}
\editors{J. van Leeuwen, ed.} 
\begin{document}

\maketitle

\begin{abstract}
 Observations of pulsar glitches remain a powerful tool for studying the interior of neutron stars. Many of the observed glitch properties are shown to result from the evolution of glitches in the different manifestations of neutron stars. Specifically, the type of glitches associated with the Crab and Vela pulsars are explained by this model. We are, also, able to adequately account for the absence, or very low rate, of glitches among the youngest and the very old pulsars.
\end{abstract}


\firstsection 
\section{Introduction}
Glitch activity, $A_g$ (which measures the mean fractional change in period per year owing to glitches), is known to vary with the characteristic age, $\tau_c$, of pulsars. We note, however, that $\tau_c$ may not give a reliable estimate of the true pulsar age in some cases. Glitch activity is a function of the size and interval (frequency) of glitches in a pulsar. Youthful pulsars with ($\tau_c \sim 10^4 - 10^5$\,yr) are known to have very high $A_g$ while younger and older ones are characterized by lower $A_g$. Urama \& Okeke (1999) observed that the glitch activity of the young and youthful pulsars generally increases with the logarithm of the spin-down rate with the exception of the Crab pulsar (younger than $10^4$\,yr), which was considered to have ``unusual'' glitch behaviours (McKenna \& Lyne, 1990). 

However, with over four decades of pulsar timing and more than 600 reported glitches (large and small) in the conventional radio pulsars, binary pulsars, millisecond pulsars, AXPs, and other manifestations of neutron stars (some of these being younger than radio pulsars), we have re-examined the relationship between some of the pulsar glitch parameters.

\section{Pulsar age, glitch size, interval and evolution}
Nearly 600 macro and micro glitches have been reported in about 130 pulsars (see e.g Chukwude \& Urama, 2010; Espinoza et al., 2011). The distribution of the glitches shows that there is a predominance of small glitches for $\tau_c > 10^5$\,yr and $\tau_c < 5 \times´ 10^3$\,yr; and no large glitch has been reported for $\tau_c > 10^7$\,yr. The preponderance of very small glitches for older pulsars may account for the non-observance of glitches for pulsars with $\tau_c > 10^8$\,yr.

To study the complex relationship between pulsar characteristic age and the size and interval of their glitches, we selected 10 pulsars with a record of at least 4 glitches of jump magnitudes $\Delta\nu/\nu \ge 10^{-8}$. Three of these pulsars are younger than $10^4$\,yr, while the rest are youthful pulsars ($\tau_c \sim 10^4 - 10^5$\,yr). 
From a plot of the average glitch interval, $\tau_g$, and a combination of glitch size and pulsar age for these pulsars, we find that the relationship between $\tau_c$, $\tau_g$ (the average glitch interval) and $\Delta\nu/\nu$ is given by:
\begin{equation}
log\, \tau_g = 0.21[log(\frac{\Delta\nu}{\nu}\tau_c)^2]+1.41[log(\frac{\Delta\nu}{\nu}\tau_c)]+4.74
\end{equation}
Here, $\Delta\nu/\nu$ is the average jump magnitude for all the glitches in a particular pulsar. $\tau_g$ is measured in days and the characteristic age, $\tau_c$, is measured in yr.
We, therefore, can estimate the average glitch size expected from pulsars of different ages. Such estimate is shown in Fig. 1.

\begin{figure}
 \vspace*{0.6 cm}
\begin{center}
 \includegraphics[width=2.8in]{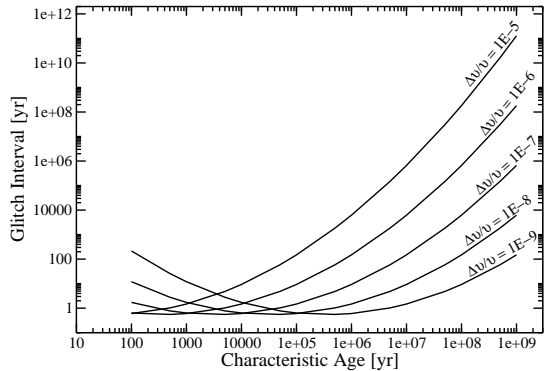} 
 \caption{Evolution of glitch sizes for different pulsar characteristic ages.}
   \label{fig2}
\end{center}
\end{figure}

Fig. 1 shows that glitches of $\Delta\nu/\nu \sim 10^{-6} - 10^{-8}$ are more frequent only in the age range $\tau_c \sim 10^3 - 10^5$\,yr. Frequent glitches of size $\Delta\nu/\nu \sim 10^{-9}$ are expected in the age range $\tau_c \sim 10^5 - 10^7$\,yr. Beyond $\tau_c \sim 10^8$\,yr, glitches of $\Delta\nu/\nu \sim 10^{-9}$ will take about 80 years to observe in a pulsar while glitch sizes greater than $\Delta\nu/\nu \sim 10^{-8}$ will take over 10,000 years to observe. This is in very good agreement with observations, as no glitch has been reported on pulsars older than $10^8$\,yr.

\section{Glitch size evolution in PSR B1737$-$30}
PSR B1737$-$30 (J1740$-$3015) belongs in the group of youthful pulsars ($\tau_c \sim 10^4 - 10^5$\,yr), known for undergoing frequent large glitches. With an average of about 1.3 glitches per year, PSR B1737$-$30 is one of the most frequently glitching pulsars of the $\sim$\,2000 known pulsars. This pulsar has a, somewhat, uniform distribution of large and small glitches. Fig. 2 shows the magnitudes and distribution of the glitches with time.

One prominent feature of Fig. 2 is a linear increase in the magnitudes of the large glitches ($\Delta\nu/\nu \ge 5 \times 10^{-7}$) observed in this pulsar. This could be an indication that different mechanisms are responsible for the large and small glitches; and that these different glitches evolve differently as shown in Fig. 1. This may actually be the best evidence of a growth in the glitch size in pulsars.

\begin{figure}
 \vspace*{0.5 cm}
\begin{center}
 \includegraphics[width=2.3in]{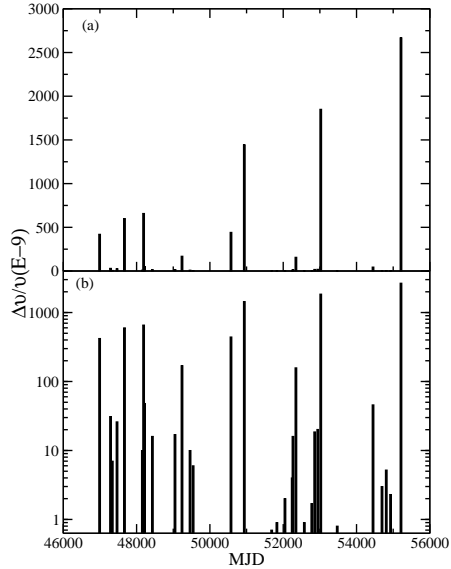} 
 \caption{Growth of glitch sizes in PSR B1737$-$30.}
   \label{fig3}
\end{center}
\end{figure}

\section{Discussions and conclusion}
Our result provides a natural explanation for the very frequent large ($\Delta\nu/\nu \sim 10^{-6}$) and super-large ($\Delta\nu/\nu \ge 10^{-5}$) glitches exhibited by very young pulsars. Frequent large and super-large glitches are expected only from pulsars in the age range $\tau_c \sim 10^2 - 10^{3.5}$\,yr. It is, therefore, not a surprise that the following youthful ($\tau_c \sim 10^4 - 10^5$\,yr) pulsars have only one super-large glitch each: CXO 1647$-$4552 ($\Delta\nu/\nu = 65 \times´10^{-6}$) and B1856+0113 ($\Delta\nu/\nu = 11.6 \times 10^{-6}$). At such ages, the frequency of such super-large glitches could range from $\sim 20 - 100$\,yr.

A growth in glitch size, similar to the one observed in PSR B1737$-30$ has also been noticed on the ``larger" ($\Delta\nu/\nu \ge 3 \times 10^{-8}$) Crab pulsar glitches. And these 3 ``larger'' Crab pulsar glitches are separated by exactly 14.5 yr (5300 d) each. If such glitch size growth continues, Crab pulsar glitch sizes of $\Delta\nu/\nu \sim 10^{-6}$ should be expected in the next 100 yr. None of the other frequently glitching radio pulsars (with a record of at least 6 glitches of sizes $\Delta\nu/\nu \ge 10^{-8}$) show such linear growth in jump magnitude. However, these frequently glitching radio pulsars are already undergoing very large glitches ($\Delta\nu/\nu \sim 10^{-6}$), implying that the glitch size may have stopped growing. However, these pulsars are younger than, or of comparable age to, PSR B1737$-$30 and, therefore, it could be difficult to understand why the glitch magnitude should still be growing in B1737$-$30. The Crab pulsar is still very young ($\tau_c = 1.2$\,kyr) of these frequently glitching radio pulsars and the glitch size had usually been explained in terms of its high interior temperature.

\section*{Acknowledgement}
JOU is very grateful for the partial support he received form IAU towards his participation in this meeting. Part of this work was done while JOU was visiting the National Centre for Radio Astrophysics of the Tata Institute of Fundamental Research (NCRA - TIFR), Pune, India. He is grateful for the C V Raman Research Fellowship that enabled him visit NCRA - TIFR.

\end{document}